# Superconducting Mg-B films by pulsed laser deposition in an *in-situ* two-step process using multi-component targets

D.H.A. Blank[*], H. Hilgenkamp, A. Brinkman, D. Mijatovic, G. Rijnders and H. Rogalla

*Low Temperature Division, Dept. of Applied Physics and MESA+ Research Institute, University of Twente, P.O. Box 217, 7500 AE Enschede, The Netherlands*

**Superconducting thin films have been prepared in a two-step *in-situ* process, using the Mg-B plasma generated by pulsed laser ablation. The target was composed of a mixture of Mg and $MgB_2$ powders to compensate for the volatility of Mg and therefore to ensure a high Mg content in the film. The films were deposited at temperatures ranging from room temperature to 300 °C followed by a low-pressure *in-situ* annealing procedure. Various substrates have been used and diverse ways to increase the Mg content into the film were applied. The films show a sharp transition in the resistance and have a zero resistance transition temperature of 22-24 K.**

Since J. Akimitsu *et al.* [1,2] reported the discovery of superconductivity in the intermetallic compound $MgB_2$, several attempts have been made to fabricate superconducting thin films. For many electronic applications and further basic studies, the availability of such films will be of great importance. To date, reports on superconducting $MgB_2$ films with the highest zero resistance transition temperatures ($T_{c,0}$), of up to 39 K, are based on a similar approach as used in wires: magnesium diffusion into boron at temperatures of 900-950 °C and at high Mg pressures [3]. Further, Eom *et al.* [4] applied this post anneal technique to room-temperature deposited Mg-B films and obtained $T_{c,0}$'s of 30-35 K. This *ex-situ* method is exploited to overcome two complicating factors for the fabrication of superconducting films of $MgB_2$: the high vapor pressure at low temperature of the magnesium and the high sensitivity of magnesium to oxidation. The first requires a post anneal step at high temperature in a high magnesium partial pressure, the latter requires very low oxygen partial pressure in the deposition system. Although these methods lead to high $T_{c,0}$ values, for a number of applications, *e.g.*, relying on multilayers and Josephson junctions, and for various basic studies it is of great importance to obtain well defined, and most favorable, epitaxial films in an all *in-situ* process, without a high-pressure post-anneal. Up to now a few attempts have been successful in obtaining *in-situ* superconducting films. To overcome the earlier mentioned complicating factors, these depositions have taken place in a two-step *in-situ* procedure [5,6]. The Oak Ridge group [6] used a thick magnesium capping layer, to avoid magnesium out-diffusion during annealing. We have used a magnesium-enriched target to obtain superconducting films on Si substrates by pulsed laser deposition (PLD), as reported in detail in Ref. [5]. In short, to avoid the loss of magnesium in the film, a Mg-enriched MgB$_2$ target has been used, prepared from a mixture of 50 Vol. % Mg powder (Alfa Aesar, purity 99.6 %) and 50 Vol. % MgB$_2$-powder (Alfa Aesar, purity 98.0 %). The targets were sintered in a nitrogen-flow for 3 hours at 640 ºC and subsequently for 10 hours at 500 ºC. The short sintering time was applied to minimize magnesium-evaporation from the pellet. After sintering, the top layer was removed to ensure that the target surface has an excess content of Mg. During deposition by PLD, special care has been taken regarding the color of the plasma. Using stoichiometric targets, as well as Mg-enriched targets or Mg metal targets, a variation in colors has been observed during ablation [5]. Depending on the argon pressure, laser fluence and ablation time, the plasma changed from green to blue. The latter plasma color is more expected, taking into account the emission spectrum of magnesium. For deposition we tuned the different deposition parameters such, that a bright blue plasma was observed. In our case, using a KrF excimer laser, the best results were obtained with an argon pressure of 0.17 mbar and a laser fluence of 4 J/cm$^2$.

---

[*] d.h.a.blank@tn.utwente.nl

The deposition temperature of the films was varied from room temperature up to 300 ºC. In first instance, Si(100) substrates were used, because of the compatibility with silicon technology. As these films show suppressed $T_{c,0}$'s, different types of substrates (SiC, because of the perfect hexagonal lattice parameter fit, with $a = 3.08$ Å and $c = 15.12$ Å, SrTiO$_3$ and MgO, the latter because of the compatibility with magnesium) were used to exclude possible negative influences of diffusion and stress, knowing that both can play a crucial role in the obtained superconducting transition temperature.

Together with the substrate, pellets of magnesium are glued on the heater. In this case, at anneal temperatures, the magnesium will evaporate causing an increase of the magnesium partial pressure during the final anneal step. Prior to the deposition, a magnesium plasma is generated by ablating from the magnesium metal target. During this procedure, care is taken to achieve the correct blue color of the plasma. This blue plasma is achieved in a rather small pressure range of about 0.17 - 0.22 mbar. Pre-ablation of the Mg as well as the Mg-enriched MgB$_2$ target is carried out to take advantage of the gettering effect of magnesium. After a few minutes, the shutter is opened and a film is deposited from the Mg-enriched MgB$_2$ target. The pressure in the chamber is adjusted to the optimal value of about 0.17 mbar and the substrates are placed in front of the target at a distance of 4.5 cm (on-axis geometry). The films are prepared at a repetition-rate of 10 Hz for 6 minutes, leading to a typical layer thickness of 200 nm.

To obtain the superconducting phase, a high-temperature annealing step is needed. This *in-situ* annealing procedure takes place in a 0.2 mbar Ar-atmosphere, at a temperature of 600 ºC. During the anneal procedure, the substrate was kept in a magnesium plasma, generated by ablating the Mg metal. The influence of annealing time and annealing temperature has been studied in more detail. From our results we conclude that the total annealing procedure has to be kept short, typically a few minutes, to avoid the Mg-evaporation out of the film. Annealing-temperatures above 600 °C did not lead to superconducting films. Furthermore, best results were obtained for those films deposited at room temperature.

As mentioned earlier, our initial results were obtained using Si substrates [5]. Figure 1 shows a typical resistance versus temperature curve for a MgB$_2$ thin film on Si (100). During the annealing procedure for this film no magnesium plasma was generated. These films exhibited a maximum $T_{c,0}$ of 16 K. This transition temperature was found to be strongly dependent on annealing time, which may be an indication of silicon diffusion. To attain higher $T_{c,0}$'s, without reverting to high-pressure Mg post-annealing, alternative substrates have been used. In Fig 2, the resistance versus temperature is given for $MgB_2$ on SiC, $SrTiO_3$, and MgO, respectively. As can be seen from the figure, the zero resistance transition temperature is in all cases 22-24 K. To our knowledge these are the highest values of films made with a low-pressure approach. The films appear to be very homogeneous; no color change has been observed over the substrate and a very sharp transition of less then one degree has been found. The films are very stable and not sensitive to moisture. The film deposited on $SrTiO_3$ shows a remarkably large RR = $\rho$ *(300 K) / $\rho$ (40 K)* of 18, which is close to the RR of 25 found for bulk material [7].

Comparable $T_{c,0}$'s of 24 K have been obtained using different substrate materials, and it seems to be a characteristic feature of the deposition procedure. To our knowledge, all superconducting magnesium-boride films reported up to now with higher transition temperatures were prepared with an *ex-situ* post-annealing step in a magnesium pressure cell [3,4]. Interestingly, all films reported up to now, made from a multi-component target and an *in-situ* anneal step [5,6] are characterized by a reduced $T_{c,0}$. Internal stress is not a likely reason for this reduced $T_{c,0}$, as similar values are obtained using different substrates and, furthermore, epitaxial growth is not expected in this two-step process.

To investigate whether magnesium-deficiency during the annealing is causing the reduced critical temperature, another procedure was applied, namely depositing a thick magnesium capping layer on top of the Mg-B film at room temperature and subsequently annealing this bilayer of Mg-B and Mg at different temperatures and times. Also sandwich structures of Mg and Mg-B have been investigated. However, no significant improvement in superconducting properties was observed.

We note that an important difference between the fabrication-processes for the Mg-annealed boron films and the PLD films from multi-compound targets is the created plasma. It is known from pulsed laser deposition that the generated plasma typically consists of different phases and that its temperature can be as high as several thousands degrees. There is a distinct difference between the substrate temperature and the temperature of the plasma. Zi-Kui Liu *et al.* [8] calculated by CALPHAD modeling the phase diagram of Mg-B under different conditions and stated that thermodynamically $MgB_2$ only can be formed under constrained conditions in terms of Mg overpressure and temperature. Apparently, multiple phases, like $MgB_4$, $MgB_7$ and solid B could be formed in the plasma and will be incorporated into the film. These phases are known to be very stable and do not dissociate into the preferred $MgB_2$ phase at the annealing temperature of 600 $^{o}$C [9].

Another possible explanation for suppressed transition temperature is impurities, like MgO. Eom *et al.* [4] showed with transmission electron microscopy that, although care was taken to avoid oxygen incorporation, a significant amount of MgO was present in the film. Especially, due to the highly energetic particles in the plasma, a partial oxidation of Mg is hard to avoid.

Furthermore, the grain size of the deposited film shows to be very small, which could be of influence on the transition temperature as well. It is noteworthy that magnetization measurements did not indicate any superconducting phases above the zero resistance transition temperature of 24 K [10].

The obtained results are of importance to grow epitaxial thin films with high $T_{c,0}$. As a consequence, using plasma techniques in combination with multi-component targets, like pulsed laser or sputter deposition, single-phase $MgB_2$ films with $T_{c,0}$'s comparable with bulk values will be exceptionally hard to realize. Alternative fabrication processes are needed to overcome these complicated factors.

In conclusion, thin films of $MgB_2$ have been fabricated on several substrates with superconducting transition temperatures up to 24 K. The deposition method was a two-step *in-situ* approach, with a deposition at reduced temperatures and a low-pressure anneal step at 600 $^o$C. Despite several optimization attempts, $T_{c,0}$ did not exceed 24 K, indicating a limitation of this deposition method. As magnesium-diffusion at high temperatures and pressures into pre-deposited boron films leads to films with superconducting properties comparable with the bulk, it is likely that the plasma temperature influences the formation of single-phase material. As a consequence, alternative deposition procedures will be required to produce epitaxially all *in-situ* grown $MgB_2$ films.

**Acknowledgements**
The authors thank C.A.J. Damen, G.J. Gerritsma, A.A. Golubov, S. Harkema, V. Leca, I. Oomen, F. Roesthuis, H.J.H. Smilde for assistance and discussions and H.M. Christen, D.C. Larbalestier, D. Schlom and X.X. Xi for valuable discussions. This work was supported by the Dutch Foundation for Research on Matter (FOM) and the Royal Dutch Academy of Arts and Sciences.

**Figures**

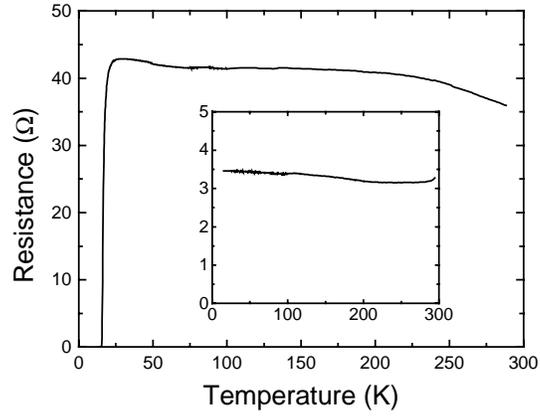

Fig.1: Resistance versus temperature for a thin film on a Si substrate. The inset shows the typical dependence for a film before the annealing.

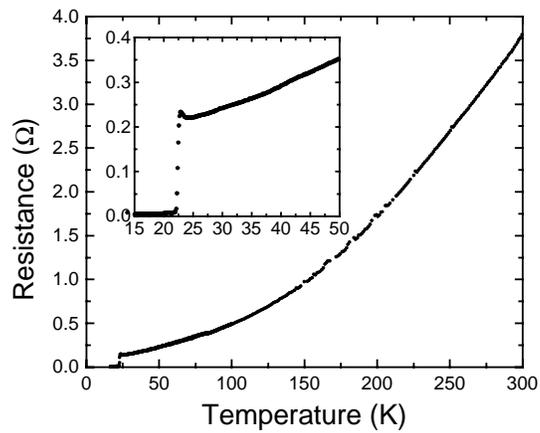

Fig. 2: Resistance versus temperature for a thin film on an SrTiO$_3$ substrate. The inset shows the transition to superconductivity on an enlarged scale.

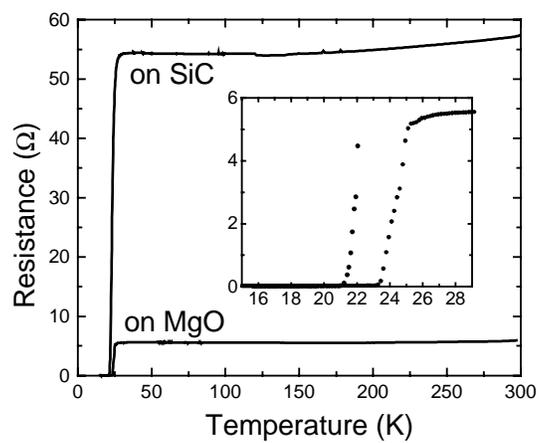

Fig. 3: Resistance versus temperature for thin films on SiC and MgO substrates. The inset shows the transition to superconductivity on an enlarged scale.